\documentclass[%
 reprint,
groupedaddress,
 amsmath,amssymb,
 aps,
]{revtex4-1}

\usepackage{graphicx}
\usepackage{dcolumn}
\usepackage{bm,color}
\usepackage[normalem]{ulem}

\usepackage{amsmath}

\begin{document}

\title{Formation and  Dynamics of Quantum Hydrodynamical Breathing Ring Solitons}

\author{Samuel N. Alperin${}^1$ and Natalia G. Berloff${}^{2,1}$}
\email[correspondence address: ]{N.G.Berloff@damtp.cam.ac.uk}
\affiliation{${}^1$Department of Applied Mathematics and Theoretical Physics, University of Cambridge, Cambridge CB3 0WA United Kingdom,\\ ${}^2$Skolkovo Institute of Science and Technology Novaya St.,
100, Skolkovo 143025, Russian Federation}


\begin{abstract}
We show that exciton-polariton condensates may exhibit a new fundamental, self-localized nonlinear excitation not seen in other quantum hydrodynamical systems, which takes the form of a dark ring shaped breather. We predict that these structures form spontaneously and remain stable  under a combination of uniform resonant and nonresonant forcing. We study single ring dynamics, ring interactions and ring turbulence, and explain how direct experimental observations might be made. We discuss the statistics of ring formation and propose an experimental scheme by which these structures may be exploited to study the smooth cross-over between equilibrium and non-equilibrium critical phase transitions. Finally, we present an alternative mechanism of  formation for these topological breathers, in which they circumscribe Gaussian resonant pumps. The observation of a breathing ring soliton would represent the first fundamental breathing soliton within the broad field of quantum hydrodynamics.
\end{abstract}

\maketitle

The spontaneous formation of patterns in disordered systems has captivated scientists for generations. Of particular interest are  self-localized patterns, which are known as solitons in integrable systems and as solitary waves in nonintegrable systems \cite{whitham2011linear,ablowitz1981solitons}. These can be understood as the fundamental excitations of nonlinear wave systems, and they are typically found in familiar forms across disparate settings. The first to be discovered were of the bright type, which are stabilized by the counterbalancing of dispersion with nonlinear effects. Other solitons are of the dark type and include vortex and domain wall type phase dislocations \cite{berry2000phase,alperin2019quantum,vilenkin1985cosmic,pismen1999vortices,pismen2006patterns,burger1999dark,swartzlander1992optical,staliunas2003transverse}.  Nonlinear optical resonators were one of the earliest physical systems in which nonlinear dynamical pattern formation was studied in depth \cite{pismen2006patterns}. The results of that field have thus served to inform the study of many other systems, such as Bose-Einstein condensates (BECs) of ultracold atoms \cite{kevrekidis2007emergent,burger1999dark} and of magnon \cite{nowik2012spatially} systems. One fundamental soliton that can be formed in the nonlinear optical resonator is the so called \textit{ring dark} or \textit{phase} soliton, formed by domain walls that close on themselves to form loops \cite{staliunas2003transverse}. In equilibrium BECs (such as ultra-cold atomic BECs), it has been shown that ring dark solitons are not stable, decaying either acoustically or into so called \textit{vortex necklaces} of more stable vortices \cite{carr2006vortices, theocharis2003ring}. These forms have been observed experimentally, though fleetingly, by trapping them along their transverse axes \cite{shomroni2009evidence}. While with this approach the ring structures are localized in that they repeatedly break down and reform within the trap, they are not self-localized and cannot exist in free space.

In this Letter we show that exciton-polariton condensates -- hybrid light-matter quantum fluids with strongly nonlinear properties and inherent nonequilibriation-- may support a different type of topological defects: \textit{breathing} ring solitons. These breathing rings are distinct from other solitary structures found in exciton-polariton condensates as well as from those found in other quantum hydrodynamical systems. Geometrically these excitations are similar to the ring shaped \textit{phase solitons} found in nonlinear optical systems \cite{taranenko1998pattern}, but are distinct in a critical way: the solitons we present here do not appear to settle into a stationary state, instead continuing to oscillate indefinitely in radius and depth \footnote{We do note that it has been predicted, in the exceptional circumstances of high nonlinearity and nonlocality, breathing optical ring -shaped solitons may exist  \cite{lu2009multiringed}. However, these solitons are of the beam type as opposed to the systems of interest here (self-localized structures within an extended system). Thus these are of a distinct class of structure.}. We study the mechanisms of formation and stabilization of these structures, and report on their dynamics and interactions, including states of ring solitonic turbulence. We also study the statistics of spontaneous ring formation, showing that breathing ring solitons can be used to study critical phenomena in systems with tunable nonequilibriation. Lastly, we demonstrate that in addition to their spontaneous formation during condensation, breathing ring solitons can be formed explicitly via localized forcing.

Condensates of exciton-polariton quasiparticles (polariton condensates) have recently been realized in semiconductor microcavities \cite{kasprzak2006bose}. As inherently nonequilibriated condensates of hybrid light-matter quasiparticles, polariton condensates may be thought of neither as equilibrium condensates or as lasers, but rather as something in between \cite{keeling2008spontaneous}. Polaritonic systems have several advantages with respect to other confined optical systems. One is  their extraordinary nonlinear properties, which arise from their excitonic component. Their  dispersion curves (the so called \textit{lower} and \textit{upper} polariton branches appearing from the hybridization of photons and excitons) allow for the individual control of the photonic and excitonic components via detuning, and their properties and dynamics can be easily accessed by angular-resolved imaging or electroluminescence spectroscopy. For low enough densities, polaritons  may be considered as bosonic quasiparticles, and so can form a coherent state  (BEC). Polariton condensates are  non-equilibrium systems  set by balance between pumping and losses due to the short lifetime of polaritons. They can continuously cross from weak coupling at higher temperatures and pumping strengths to strong coupling at lower temperatures and lower pumping intensities. In addition, the lifetime of polaritons  in the microcavities systems can be increased by improving the quality and number of dielectric Bragg mirrors. Therefore, depending on the polariton lifetime and pumping intensities polariton condensates continuously cross between strongly non-equilibrium systems -- lasers-- (short lifetime and large pumping rates) and equilibrium BECs (large lifetimes and small pumping). 

Polariton condensates can be excited by two different types of pumping. In the resonant pumping scheme, the energy and angle of incidence of the excitation laser are set to be quasi-resonant with a mode of the lower polariton branch. In the non-resonant pumping scheme, the excitation laser has an energy much higher than that of the lower polariton branch. In this case, polaritons can spontaneously form macroscopic coherent states -- polariton lasers or polariton condensates-- by the accumulation of particles in the same quantum state \cite{kasprzak2006bose}. Recent experiments have started to combine resonant and  non-resonant pumping   \cite{ohadi2016tunable}.  In those experiments, chemical etching of a GaAs substrate allowed resonant excitation from the back side of the cavity, preventing backscatter from the non-resonant pumping, while allowing synchronization. This technique makes it possible to independently vary the pumping intensity distributions of resonant and  non-resonant excitations.

The mean-field behaviour of polariton condensates is governed  by the generalized  complex Ginzburg-Landau equation  (cGLE), with the condensate wavefunction $\psi({\bf r},t)$   coupled to the hot exciton reservoir density $N_R$ \cite{kalinin2018simulating,keeling2008spontaneous, wouters2007excitations, carusotto2013quantum,keeling2011exciton}, so that 

\begin{eqnarray}
  \label{GPE}
 i \partial_t\psi &= &
 -(1-i\eta N_R) \nabla^2 \psi
    + |\psi|^2 \psi+g N_R \psi \\ \nonumber &+& i(N_{R}-\gamma)\psi+V_{\rm ext}\psi+i \bar{P}\psi^{*(n-1)}\\
 \partial_t N_R&=&P-(b_0+b_1|\psi|^2)N_R. \label{NR}
 \end{eqnarray}
 where we set $\hbar=1$ and $m=1/2$. In these coupled  equations,  $g$ is proportional to the polariton-exciton interaction strength, $\eta$ represents the energy relaxation \cite{wouters2012energy,berloff2013universality}, $\gamma$  and $b_0$ are proportional to the inverse lifetime of the polariton  and hot excitons, respectively,   $b_1$ is proportional to the ratio of the interaction strength between the condensate and the hot excitons to that  between condensate particles. The incoherent and resonant (at $n:1$ resonance with the condensate frequency) pump sources are described by  the pumping intensities $P({\bf r},t)$ and $\bar{P}({\bf r},t)$, respectively \footnote{The fixed parameter values in our simulations are $b_0=1$, $b_1=1$, $\gamma=0.3$, and $\eta=0.3$}. The sample disorder is described by a spatially dependent external potential $V_{\rm ext}({\bf r})$. 
 The steady state of the systems is  characterized by the chemical potential $\mu$. Using the Madelung transformation $\psi=\sqrt{\rho({\bf r})} \exp[i S({\bf r})+i\mu t]$, neglecting the energy relaxation and the external potential in the view of their smallness and replacing the steady state of Eq.~(\ref{NR}) $N_R= P/(b_0 + b_1\rho)$ with its Taylor expansion $N_R\approx P/b_0-Pb_1\rho/b_0^2$, we transform Eqs. (\ref{GPE}-\ref{NR}) into the continuity equation and the integrated form of the Bernoulli equation (with ${\bf u}=\nabla S$) as
\begin{eqnarray}
\tilde{\mu}&=&\biggl(1-\frac{gPb_1}{b_0^2}\biggr)\rho +u^2 -\frac{\nabla^2 \sqrt{\rho}}{\sqrt{\rho}}-\bar{P}\sin2S, \label{mu}\\
\nabla\cdot(\rho {\bf u})&=&\biggl(\frac{P}{b_0}-\frac{Pb_1}{b_0^2}\rho-\gamma + \bar{P} \cos2S\biggr)\rho \label{cont}
\end{eqnarray}
with the solution $S=0, \rho=(P b_0 -\gamma b_0^2+\bar{P}b_0^2)/Pb_1$, $\tilde{\mu}=(1-gPb_1/b_0^2)(P b_0-\gamma b_0^2+\bar{P}b_0^2)/Pb_1,$ and $\mu=\tilde{\mu}+gP/b_0$. Nontrivial uniform state is formed when $\gamma b_0< P + \bar{P} b_0$.

\begin{figure}[!t]
\centering
\includegraphics[width=\columnwidth]{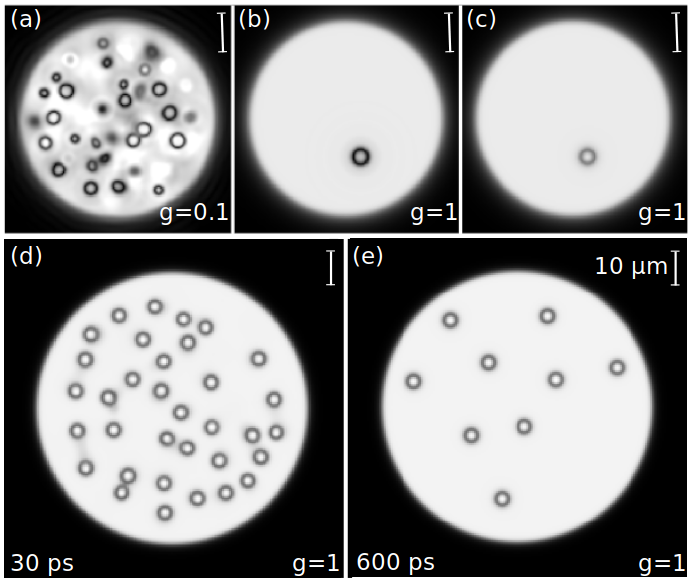}
 \caption{Spontaneously formed breathing rings in  exciton-polariton condensates for $P=\bar{P}=5$. Density contourplots of the condensate shown illustrate (a) a time- snapshot of ring turbulence; (b)  a quasistationary state with a single ring; (c)  time-averaging of (b) over many ring oscillations; (d-e) different stages of the condensate evolution  averaged over the time-scale of the ring oscillation. }
  \label{figure1}
\end{figure}

To form ring solitons on the top of this nontrivial uniform state, we must enter the regime in which Ising-type phase domains are supported. This regime is that of phase bistability, which can be achieved in the complex Ginzburg-Landau equation via parametric forcing induced by a relatively large second order resonant pump \cite{coullet1992strong}. We thus set the resonant forcing term to $n=2$ in Eq.~(\ref{GPE}).  This regime surrounds a fork-type Hopf bifurcation near the generation threshold as opposed to a sub or supercritical one \cite{staliunas2003transverse}. Assuming that the condensation process arises from a noisy initial state, the condensate forms with independently coherent regions separated by domain walls of either Bloch or Ising type. 
The healing length of the domain wall $L_h$ -- the characteristic distance of the density recovery when locally perturbed --  is defined by the length scale at which the potential and kinetic energies are equal, which by Eq. (\ref{mu}-\ref{cont}) is   $L_h\approx |\mu|^{-1/2}$ defined above. In typical GaAs experiments the  reservoir relaxation times are large so increasing $P(\bar{P})$ or/and $g$ is  the most natural way to decrease the  healing length of the domain wall.

Ising walls are required to either end at the boundary of the condensate, or to form closed loops. In the latter case, the dark ring grows or shrinks until it reaches a characteristic radius \cite{staliunas2003transverse}.  In the optical case the ring of characteristic radius finds itself in a stable, local energy minima, and thus remains stationary.  However, in polariton condensates, after reaching the critical radius the behavior of the rings diverges from the behavior of geometrically similar solitons observed in other systems. We determined that the ring solitons self-annihilates   (i) in the fast reservoir regime $b_0\gg \gamma$, (ii)  in small reservoir detuning regime $g\ll1$, (iii)  in long-lived polaritonic systems. All these regimes are physically relevant to some experiments  \cite{nelsen2013dissipationless,kalinin2019polaritonic,berloff2017realizing}. However,    a slow reservoir evolution ($b_0\lesssim\gamma$), for short-lived polaritons and a sufficiently large reservoir detuning (all of which correspond to values of typical GaAs microcavity experiments \cite{kasprzak2006bose,deng2007spatial,manni2012penrose}) prevent the ring soliton from disappearing and lead to appearance of a ring breather: the dissipative decrease in the radius of the ring soliton is accompanied by the increase in the reservoir profile density in the ring core, which imposes a repulsive force in the outward direction to make the ring expand. This process repeats itself as shown in Supplemental Video 1. This nonlinear excitation is self-localized by an explicitly dynamical interaction.

As all of the destabilizing mechanisms (i)-(iii) have similar effects on the existence and dynamics of the ring solitons, we concentrate on the effect of varying the polariton-exciton interaction strength (parameterized by $g$). The detuning between the cavity
photon energy and the exciton resonance determines the relative photonic/excitonic character of the polariton and, therefore, its effective mass and the strength of the  polariton-exciton
interactions  \cite{carusotto2013quantum}. The detuning $g$ can be further changed by the pumping geometry by  considering trapped
condensates separated from the pumps \cite{cristofolini2013optical}. Finally, implanting protons into the quantum wells or into the top of distributed Bragg reflectors allows for an independent spatial control of both
the exciton and the cavity photon energies, and, therefore, affects $g$ \cite{schneider2016exciton} as well. By these mechanisms, the experimental ranges of our dimensionless parameter $g$ can vary between $0.1 - 2$. 

Our simulations show that in the case of low $g$, rings form ad infinitum. This makes for a sustained state of ring turbulence. A condensate density for a particular moment of time  in this regime is shown in  Fig. \ref{figure1}(a). While the dynamics  of the low $g$ case are not uninteresting from a theoretical perspective, experiments on exciton-polariton condensates are currently limited to time-averaged imaging, and thus this case is not directly observable. However we find that in the high $g$ case, rings are formed only during the condensation process. They appear to interact attractively, and upon contact a pair of rings will either merge into one or annihilate each other. Eventually the decay of rings ends, and a quasistationary state is reached with rings being pinned by the system disorder represented in our simulations by $V_{\rm ext}$. This is demonstrated in Supplemental Video 2, time snapshots from which are shown in Fig. \ref{figure1}(d,e) illustrating the condensate density contour plots. Figure \ref{figure1} shows a time snapshot of a spontaneously formed breathing ring soliton after the system has reached its final, quasistationary state (b), as well as a time integrated image of that state (c). 
Thus, we predict that long-lived breathing ring solitons are directly observable, and that their ring shaped character, radii, locations and numbers are directly measurable as well.

\begin{figure}[]
\centering
\includegraphics[width=\columnwidth]{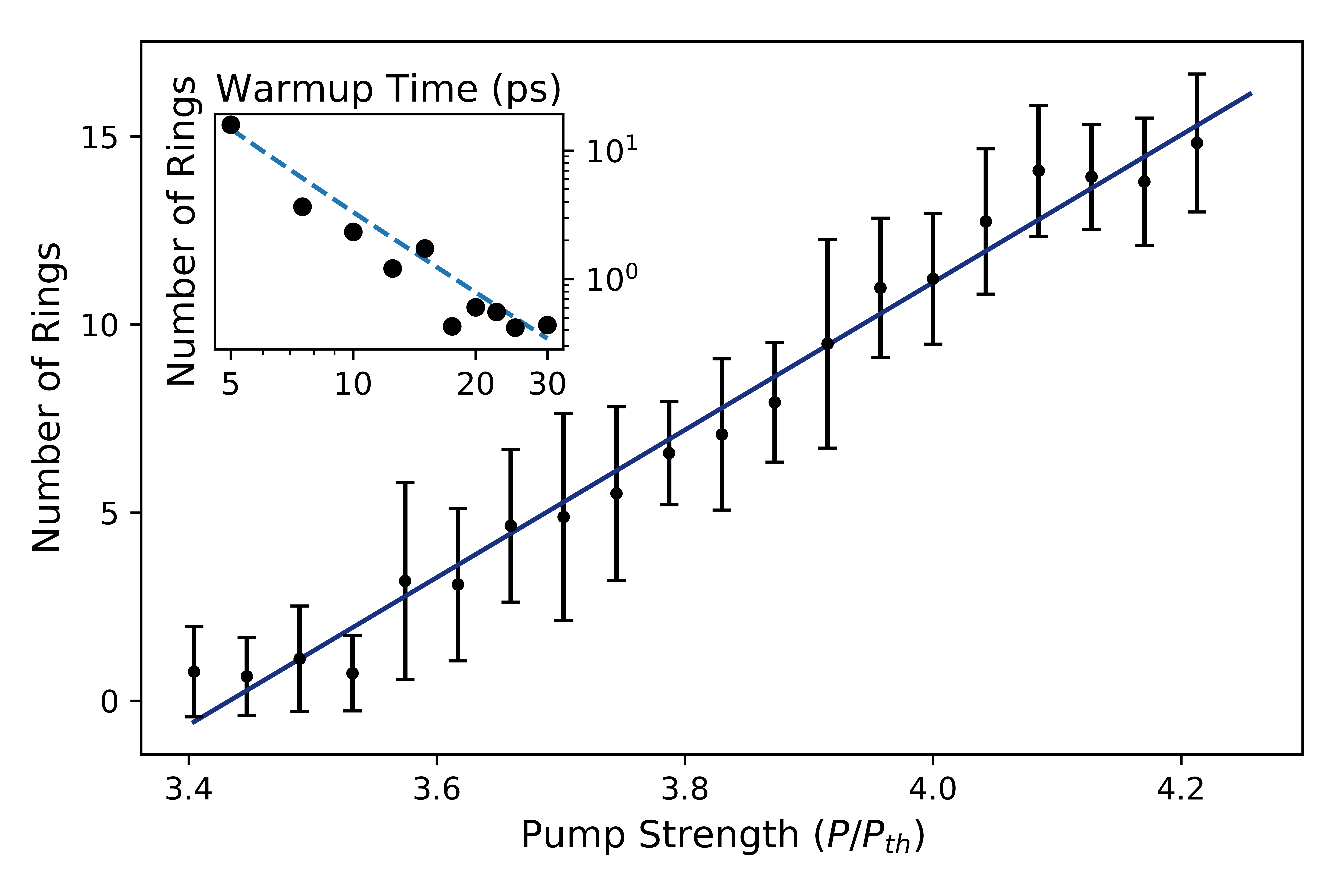}
 \caption{Number of rings in the quasistationary state as a function of pump strength ($P=\bar{P}={\rm const}$), in units of the threshold pump strength $P_{\rm th}$. Results are averaged over $10$ random iterations of initial noise and potential disorder. A linear fit is shown in blue. The inset shows a log-log plot of the number of rings in the quasistationary state as a function of warmup-time, defined as the time over which the pumps are increased to a fixed amplitude ($P=\bar{P}=4.2P_{th}$). A dashed blue line shows the power law $t^{-2}$. }
 \label{figure2}
\end{figure}

The mechanism by which breathing ring solitons have been shown to form for high $g$ resembles Kibble-Zurek (KZ) mechanism of defect formation in equilibrium systems \cite{damski2010soliton,zurek1996cosmological}. The KZ mechanism was first understood in the context of the  phase transitions in the early Universe \cite{kibble1976topology,kibble1980some, zurek1985cosmological}, and later in liquid 4He and 3He, liquid crystals, superconductors \cite{bauerle1996laboratory,ruutu1996vortex,bowick1994cosmological,maniv2003observation}, equilibrium Bose-Einstein condensates \cite{berloff2002scenario,weiler2008spontaneous}.  The  similarities and differences between the KZ transition and pattern formation in nonequilibrium systems are the subject of intense exploration, with an emphasis on the common mechanism of the defect formation: locally uniform symmetry breaking in separate parts of the system which cannot communicate in a finite time, and which thus form to be globally nonuniform to a degree set by the speed of the phase transition (the quench rate). The  main difference between the KZ transition and pattern forming in nonequilibrium systems is that in the former, it is assumed that the system  is driven out of equilibrium only in the vicinity of the phase transition \cite{zurek1985cosmological}.  In spite of extensive research on both the KZ transition and on pattern formation in a wide variety of nonequilibrium systems, questions remain  regarding the nature of the cross-over between the two mechanisms, and regarding the types of the defects that they can result in. We investigate this relationship by counting the number of quasistationary rings formed spontaneously from random initial noise in the presence of a small sample disorder. This disorder does not hamper the formation of rings, but rather acts like sandpaper, resisting their movement across its surface. Fig. \ref{figure2} shows the resulting  linear, positive correlation between pump power and ring soliton density.
To elucidate the effect of the quenching time on the defect formation, we repeated our simulations linearly increasing the pumps from zero to $P=\bar{P}$ over different timescales. The results, shown in the inset of Fig. \ref{figure2}, reveals  a $k=-2$ power law. We note that recent theoretical work on nonequilibrium holographic superfluids have shown qualitatively similar results: a linear dependence of excitation strength (temperature in that context) on defect (vortex) density, and a power law dependence of quench time on defect density \cite{chesler2015defect}.

\begin{figure}[!t]
\centering
\includegraphics[width=\columnwidth]{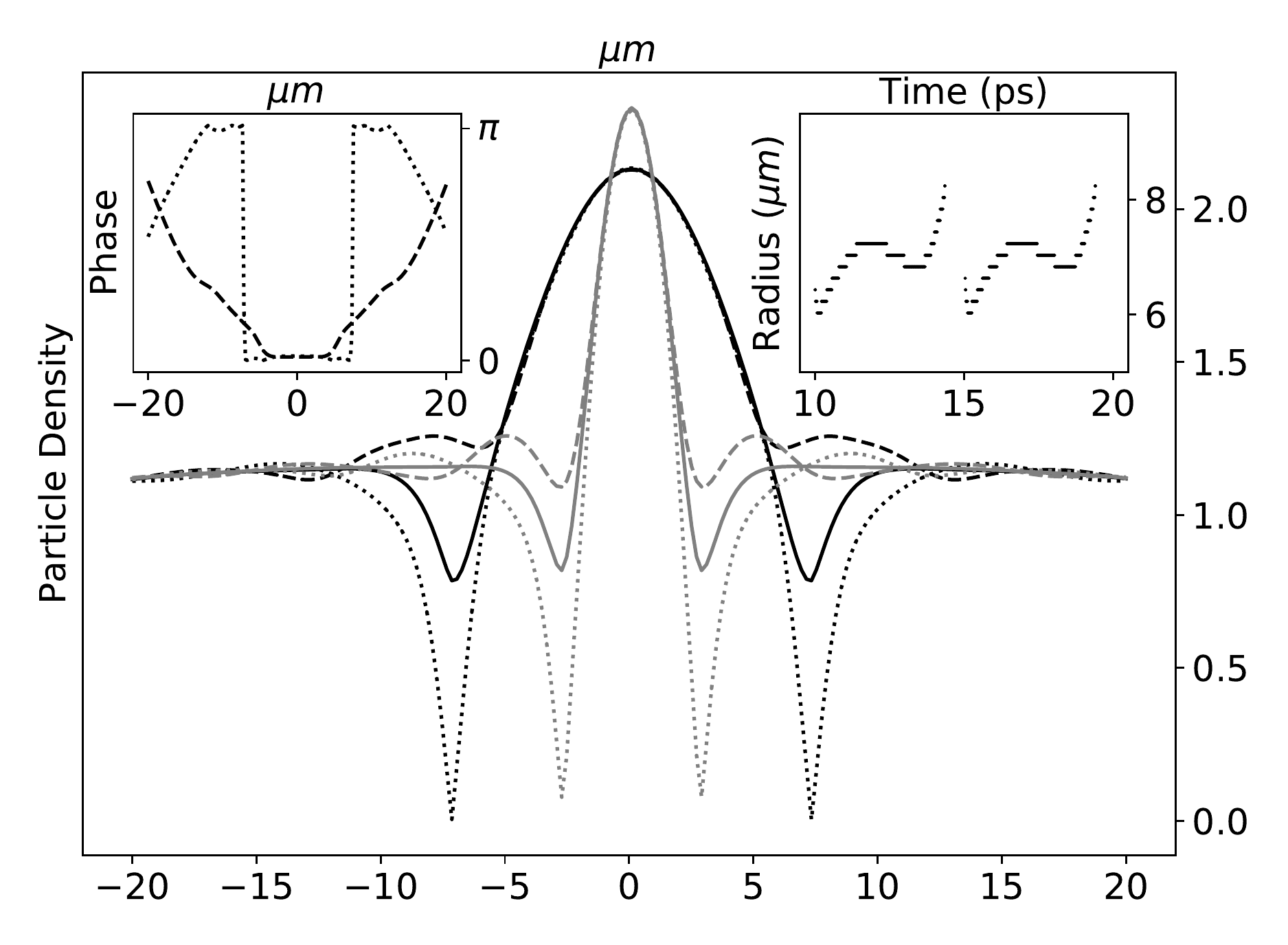}
 \caption{Central density cross sections of a ring soliton formed explicitly by the first-order resonant Gaussian excitation of a uniform, nonresonantly excited condensate. The profiles are shown for two resonant pumps of different width (black and gray). These profiles are shown at multiple times over the course of a single oscillation (dotted and dashed), along with time-averaged profiles (solid). In the left inset, the phase structure of the ring  are shown. The right inset shows the radius of the breathing ring over time, over two periods. The curve is discontinuous, demonstrating that the ring regularly collapses and reforms. Simulation parameters are as follows: $\beta=0.05$, $P=1$, $A_{\rm res}=10$ (black lines) and $\beta=0.4$, $P=1$, $A_{\rm res}=14.1$ (grey lines).}
  \label{figure3}
\end{figure}

In addition to the their spontaneous nucleation during the condensate formation, breathing ring solitons can be formed  explicitly, by either first or second order resonant forcing  with a non-uniform (e.g. Gaussian) profile. With direct numerical integration we simulate the effect of an exciton-polariton condensate excited by a uniform nonresonant pump of uniform amplitude $P$ combined with a small resonant pump with the Gaussian profile $\bar{P}=A_{\rm res}\exp^{-\beta r^2}$ where $\beta$ is an inverse width parameter. Either at first ($n=1$)  or second  ($n=2$) order resonant frequency, the resonant pump locally breaks gauge symmetry and fixes the phase of the condensate \cite{burke2008classification}. The surrounding condensate remains gauge symmetric, however, and the phase along the ring shaped interface between these regions cycles regularly; when the phase difference between the interfacial ring and the resonant spot is $\pi$, the ring is momentarily singular. Unlike spontaneously formed rings, these rings may be also formed under the fast-reservoir approximation and across wider range of parameters. Figure \ref{figure3} shows cross sections of rings formed by localized resonant excitation for two different resonant pumping profiles (black and gray). We plot the time-averaged profiles (solid lines) as well as two different time-snapshots of the intensity profiles. The ring radius over time is plotted for one of these simulations (right inset) over two periods. The discontinuities of the ring radius illustrate that the ring does not survive continuously, but instead regularly explodes and reforms from outside of the interfacial region. The null density of the ring minimum corresponds to phase discontinuity, indicative of a topological defect.

In conclusion, we have theoretically predicted the spontaneous formation of stable breathing ring solitons in exciton-polariton condensates. The proposed experimental realisation for such novel topological defects is  well within  the current  experimental conditions and properties of existing microcavities. 
These structures would represent the first fundamental breather and the first stable ring soliton in any quantum hydrodynamical system. We have discussed how the stability of breathing ring solitons can be exploited to study nonequilibrium defect formation statistics, and thus to probe the fundamentals of nonequilibrium phase transitions. Further, we have proposed an experimental scheme by which these statistics could be probed over the continuous crossover between equilibrium and nonequilibrium phase transitions. Finally, we have elucidated how ring solitons can be formed explicitly, via localized resonant excitation of either second or first order resonant pumping frequency. 

{\it Aknowledgements.} NGB acknowledges the support from Skoltech-MIT NGP. Both authors are grateful to Kirill Kalinin for discussions.

\end{document}